\documentclass[10pt,onecolumn,aps,prd,preprintnumbers,showpacs,superscriptaddress,nofootinbib,amsmath,amssymb,floats,floatfix,showkeys,notitlepage,longbibliography]{revtex4-1}

\usepackage{orcidlink}
\usepackage{comment}
\usepackage{lipsum}
\usepackage{graphicx}
\usepackage{subfigure}
\usepackage{palatino}
\usepackage{sans}
\usepackage{hyperref}
\hypersetup{colorlinks=true,linkcolor=blue,urlcolor=blue,citecolor=blue}
\usepackage[toc,page]{appendix}
\usepackage[normalem]{ulem}
\usepackage{adjustbox}
\usepackage{latexsym}
\usepackage{amsmath}
\usepackage{amssymb}
\usepackage{amsfonts}
\usepackage{dcolumn}
\usepackage{bm}
\usepackage{tikz}
\usepackage{bigints}
\usepackage{array,tabularx,multirow,booktabs}
\usepackage[tracking=true]{microtype}
\usepackage{soul} 
\usepackage{booktabs}
\usepackage{multirow}
\usepackage{braket}
\SetTracking{}{500}
\SetTracking{encoding={*}, shape=sc}{40}
\UseRawInputEncoding 
\allowdisplaybreaks

\begin{document} \sloppy
\title{HBAR entropy 
 of Infalling Atoms into a GUP-corrected Schwarzschild Black Hole and equivalence principle}

\author{Ali \"Ovg\"un \orcidlink{0000-0002-9889-342X}}
\email{ali.ovgun@emu.edu.tr}
\affiliation{Physics Department, Faculty of Arts and Sciences, Eastern Mediterranean University, Famagusta, 99628 North
Cyprus via Mersin 10, Turkiye.}

\author{Reggie C. Pantig \orcidlink{0000-0002-3101-8591}} 
\email{rcpantig@mapua.edu.ph}
\affiliation{Physics Department, School of Foundational Studies and Education, Map\'ua University, 658 Muralla St., Intramuros, Manila 1002, Philippines.}

\begin{abstract}
In this work, we have investigated the phenomenon of acceleration radiation exhibited by a two-level atom freely falling into a Generalized Uncertainty Principle (GUP)-corrected Schwarzschild black hole. We derive analytic expressions for the atom's excitation probability with simultaneous emission of a scalar quantum and observe that it satisfies the Einstein equivalence principle when compared to the excitation probability induced by a uniformly accelerating mirror, motivated by studies  [10.1103/PhysRevLett.121.071301] and [10.1073/pnas.1807703115]. 
Adopting an open-quantum-system framework, we then compute the horizon-brightened acceleration radiation (HBAR) entropy for the GUP-corrected spacetime and find that it reproduces the Bekenstein-Hawking entropy law, with corrections characteristic of GUP effects. These results underline the robustness of thermal radiation processes near horizons and the universality of entropy corrections in quantum-improved black hole spacetimes.
\end{abstract}


\date{\today}
\maketitle


\section{Introduction} \label{intro}

General relativity unified spacetime geometry with gravitation \cite{Einstein:1915ca}. Its merger with thermodynamics and quantum field theory then revealed profound effects: black hole entropy \cite{Bekenstein:1973ur,Bekenstein:1974ax,Bekenstein:1980jp}, Hawking radiation \cite{Hawking:1975vcx,Hawking:1976de,Hawking:1976ra}, particle emission \cite{Page:1976df,Page:1976ki,Page:1977um} as well as the Fulling-Davies-Unruh effect \cite{Fulling:1972md,Davies:1974th,Unruh:1976db,Unruh:1983ms,Muller:1997rt,Vanzella:2001ec,Crispino:2007eb}. More recently, optical analogues such as ultraslow light experiments \cite{weiss2000black}, fiber-optic event-horizon simulations \cite{Philbin:2007ji}, and entanglement-entropy studies \cite{Das:2005ah} have further illuminated these connections. Scully and colleagues later showed that acceleration radiation can be viewed as real photons arising from interrupted atomic virtual transitions \cite{Scully:2003zz,Belyanin_2006}. Then Scully et al. extended this idea to atoms free-falling through the Boulware vacuum into a black hole \cite{Boulware:1974dm,Scully:2017utk}: although their proper acceleration vanishes, their relative acceleration with respect to the external field modes gives rise to observable acceleration radiation.

The thermodynamic interpretation of black holes stands as one of the most profound insights in theoretical physics, revealing a deep interplay between gravity, quantum theory, and statistical mechanics \cite{Bardeen:1973gs,Hawking:1975vcx,Kempf:1994su,Ashoorioon:2004vm,Das:2011tq}. The seminal works by Bekenstein and Hawking established that black holes possess entropy proportional to the area of their event horizons, formally expressed as 
$S_{\rm BH} = kc^3A / 4\hbar G$,
thereby aligning black hole physics with the laws of thermodynamics \cite{Bekenstein:1973ur, Hawking:1976de}. This entropy-area relationship not only suggests that black holes have a micro-physical basis, potentially statistical in origin, but also challenges classical intuitions by implying a thermodynamic temperature associated with spacetime geometry \cite{Unruh:1976db,Gibbons:1977mu}. The thermodynamic formalism has further prompted a re-examination of entropy's statistical origin in gravitational systems, raising the possibility that black hole entropy may emerge from microscopic quantum degrees of freedom that are yet to be fully understood \cite{Bombelli:1986rw, Srednicki:1993im}. These results have been instrumental in advancing quantum gravity proposals, particularly string theory and loop quantum gravity, which aim to provide microstate descriptions consistent with the Bekenstein-Hawking entropy formula \cite{Strominger:1996sh, Ashtekar:1997yu}. \textcolor{black}{Recent developments further explore area-law entanglement in dynamical contexts, such as cosmological expansions, gravitational collapse, regular black holes, and field-curvature couplings, revealing deviations from the strict area law and implications for entropy in quantum gravity \cite{Belfiglio:2025cst,Belfiglio:2025hzo,Belfiglio:2024wel,Belfiglio:2024qsa,Belfiglio:2023sru}.} In this framework, entropy no longer merely quantifies disorder, but serves as a key probe into the fundamental architecture of spacetime and quantum fields, highlighting black holes as natural laboratories for exploring the synthesis of general relativity and quantum mechanics \cite{Jacobson:1995ab,Carlip2000}. Researchers have explored various alternative interpretations and refinements of the Equivalence Principle. In Ref. \cite{Singleton:2011vh}, Singleton and Wilburn compare Hawking and Unruh radiation, showing how their differing detector responses in curved and accelerated frames probe the limits of the equivalence principle. In \cite{Singleton:2016yal}, Singleton and Wilburn contrast the global formulation of Mach's principle with the local equivalence principle, highlighting fundamental differences between inertial frames determined by distant matter and those defined by local gravitational experiments.  On the other hand, in \cite{Demir:2022jvc}, Demir investigates the scattering times of quantum particles in a gravitational potential and identifies signatures of potential violations of the equivalence principle. In \cite{Scully:2017utk}, Scully et al. develop a quantum-optics framework to describe radiation emitted by atoms freely falling into a black hole.
Camblong et al. analyze the near-horizon conformal symmetry underlying acceleration radiation during atomic free fall into a Schwarzschild black hole \cite{Camblong:2020pme}. Azizi et al. extend this analysis to Kerr black holes, revealing how near-horizon conformal quantum mechanics governs atomic acceleration radiation \cite{Azizi:2020gff}. In \cite{Azizi:2021qcu}, the master equation for acceleration radiation is derived via a conformal quantum-mechanics mapping. The thermodynamic implications of this acceleration radiation are explored through conformal quantum mechanics in \cite{Azizi:2021yto}. Sen et al. investigate quantum corrections to the equivalence principle and compute the associated HBAR entropy \cite{Sen:2022tru}. The near-horizon structure of acceleration radiation across a class of static, spherically symmetric black hole geometries is studied in \cite{Sen:2022cdx}.
In \cite{Sen:2023zfq}, the spectrum of acceleration radiation for a two-level atom falling into a Kerr-Newman black hole is examined beyond the near-horizon approximation. Das et al. characterize HBAR in braneworld black hole spacetimes, highlighting brane-induced modifications \cite{Das:2023rwg}. Kaul and Majumdar derive logarithmic quantum corrections to the Bekenstein-Hawking entropy, refining the entropy-area relationship for black holes \cite{Kaul:2000kf}. Jana et al. quantitatively analyze the HBAR entropy of an atom falling into a quantum-corrected charged black hole, extending prior semi-classical treatments \cite{Jana:2024fhx}. In \cite{Rahaman:2025mrr,Rahaman:2025grm}, Rahaman analyzes the behavior of an atom placed in the vicinity of a Lorentz-violating Kalb-Ramond black hole background, uncovering novel signatures of Lorentz violation in the atom's radiation response. In \cite{Jana:2025hfl}, Jana et al. derive an inverse logarithmic correction to the HBAR entropy of an atom falling into a renormalization-group-improved charged black hole.

One of the key theoretical predictions emerging from various approaches to quantum gravity is the existence of a minimum measurable length, which modifies the Heisenberg uncertainty principle at Planck-scale energies \cite{Amati:1988tn,Garay:1994en,Kempf:1994su,Ashoorioon:2004vm,Das:2011tq}. This idea is formalized through the GUP, which alters the standard commutation relations between position and momentum operators to include quadratic (and sometimes linear) corrections in momentum, thereby encoding a minimal length scale into the fabric of spacetime \cite{Maggiore:1993rv,Kempf:1994su}. Such modifications have been derived not only from string theory and loop quantum gravity but also from black hole \textcolor{black}{gedanken} experiments, where localization of particles near the Planck scale leads to gravitational back-reaction effects that limit spatial resolution \cite{Scardigli:1999jh,Hossenfelder:2012jw}. In the presence of a GUP, black hole thermodynamics undergoes significant modifications: the black hole temperature and entropy receive quantum corrections, potentially halting evaporation at a finite remnant mass and resolving divergences near the end of the evaporation process \cite{Adler:2001vs, Cartin:2004cb}. The impact of the GUP on the quasinormal modes and shadow of the Schwarzschild black hole is investigated in detail in Ref.~\cite{Anacleto:2021qoe}. Quantum corrections to the scattering and absorption cross-sections of a Schwarzschild black hole arising from GUP are comprehensively analyzed in Ref.~\cite{Anacleto:2020lel}. The effects of GUP on the scattering and absorption phenomena for \textcolor{black}{extra-dimensional} black holes are explored in Ref.~\cite{Anacleto:2023ntm}. A notable consequence is the correction to the black hole entropy-area relation, where the Bekenstein-Hawking formula receives logarithmic and inverse area terms, consistent with predictions from other quantum gravity models \cite{Medved:2004yu,Nozari:2008rc}. These corrections are not mere formal modifications, but are expected to play a crucial role in addressing the black hole information paradox and understanding the endpoint of black hole evaporation \cite{Chen:2002tu,Nozari:2006gg}. Therefore, the GUP framework has become an effective tool for probing Planck-scale physics in a semi-classical setting, allowing theoretical predictions to remain within reach of future quantum gravity phenomenology or black hole analog models \cite{Das:2008kaa,Ali:2009zq,Scardigli:1999jh, Adler:2001vs}. 
In the pursuit of reconciling gravitational dynamics with quantum information theory, one promising avenue involves analyzing the entropy flow associated with quantum systems interacting with black holes, particularly when modeled as infalling matter \cite{Bekenstein:1980jp, Unruh:1982ic}. A particularly insightful approach to black hole entropy generation is based on analyzing the entropic impact of atomic systems interacting with quantum fields near the horizon, such as those modeled as two-level systems coupled to vacuum fluctuations \cite{Scully:2003nwl,Alsing:2005dno,Svidzinsky:2018jkp}. Within this context, the concept of HBAR entropy has emerged as a measure of entropy increase associated with quantum atoms falling into black holes, particularly when entropy is defined through changes in the reduced density matrix of quantum fields. This entropy, referred to as HBAR entropy, conceptually represents the quantum-statistical entropy extracted from field correlations and energy levels of a matter system undergoing interaction with curved spacetime geometry. While HBAR entropy is not a widely standardized term in mainstream literature, it is likely to be rooted in quantum information thermodynamics under relativistic conditions. Comparable models include atomic detectors in Unruh-Dewitt setups, where entropy is calculated from field entanglement dynamics. Studies modeling the fall of two-level atoms into Schwarzschild backgrounds have demonstrated that the interaction between the atom and the near-horizon field leads to a nontrivial entropy increase, interpreted as radiation entropy, even when the black hole remains semiclassical \cite{Scully:2017utk, Hu:2008rga,Svidzinsky:2018jkp}. The HBAR entropy model captures the net entropy flux due to vacuum polarization and excitation-deexcitation processes of quantum matter as it traverses the strong gravitational gradients near the event horizon, enabling a thermodynamic description of black hole-field interaction \cite{Page:1993df, Brustein:2013qma}. Incorporating GUP effects into this framework further modifies the entropy transfer dynamics, since the atom-field interaction now unfolds in a quantum-corrected spacetime geometry, introducing corrections to the detector's transition rates and entropy exchange \cite{Nozari:2012nf, Chen:2014xsa}. Thus, HBAR entropy provides a localized probe into Planck-scale entropy generation driven by quantum measurement processes, serving as a micro-physical diagnostic for quantum gravity-induced corrections to black hole thermodynamics.

The study of atoms or elementary quantum systems falling into black holes has become a powerful model to probe the nature of entropy production, decoherence, and information dynamics near horizons \cite{Bekenstein:1974ax,Unruh:1982ic}. A paradigmatic insight from Bekenstein's \textcolor{black}{gedanken} experiments was that the act of dropping quantum matter into a black hole can yield quantifiable increases in black hole entropy, constrained by the Generalized Second Law (GSL) of thermodynamics \cite{Bekenstein:1980jp,Sorkin:1986mg}. In these models, quantum atoms-frequently modeled as two-level systems or Unruh-DeWitt detectors-interact with the quantum vacuum near the horizon, experiencing field excitations due to curvature, acceleration, or horizon-crossing effects \cite{Unruh:1976db, Takagi:1986kn}. As these atoms fall into the black hole, they undergo irreversible decoherence and thermalization, effectively transferring information and entropy into the field modes and black hole interior, leading to a net increase in total entropy as required by the GSL \cite{Anglin:1995pg, Lombardo:1995fg}. These interactions provide a natural arena to study the quantum information loss paradox, as the process appears to convert pure quantum states into mixed thermal states, violating unitarity unless some mechanism preserves or encodes the information \cite{Hawking:1976ra, Preskill:1992tc}. Moreover, models of infalling atoms can be used to calculate entropy production rates in quantum field theory in curved spacetime, offering quantitative insight into how information disperses across event horizons through quantum channel-like behavior \cite{Martin-Martinez:2012chf, VerSteeg:2007xs}. \textcolor{black}{Related advances in quantum communication near black holes, including field-mediated channels and cryptographic interpretations of Hawking radiation decoding, highlight how curved spacetimes influence information transfer and security \cite{Jonsson:2020npo,Brakerski:2022zcx}.} Importantly, these models are also consistent with open quantum systems approaches, where the atom acts as a probe interacting with an environmental field, and the resulting entropy change reflects the loss of information into inaccessible degrees of freedom such as those behind the horizon \cite{Breuer:2007juk, Benatti:2005uq}. When such infalling atomic systems are analyzed under GUP-modified spacetime, new corrections appear in transition probabilities, proper time evolution, and field commutation structures, ultimately affecting the entropy flow and possibly offering a route to information recovery mechanisms \cite{Das:2008kaa,Ali:2009zq,Abutaleb:2013vxa,Deb:2016psq,Das:2020ujn}. Thus, infalling atoms serve as ideal quantum thermodynamic probes for detecting Planck-scale effects in black hole environments, bridging the domains of field theory, information dynamics, and semiclassical gravity. \textcolor{black}{The novelty of this work lies in applying the HBAR framework, where entropy arises from infalling atoms as quantum probes, to a GUP-corrected spacetime, deriving explicit corrections to excitation probabilities and entropy that match those from other quantum gravity approaches \cite{Kaul:2000kf,Medved:2004yu,Gour:2003jj,Chatterjee:2003uv}. Unlike prior HBAR studies on classical \cite{Scully:2017utk,Svidzinsky:2018jkp} or charged black holes \cite{Jana:2024fhx}, we verify the robustness of the equivalence principle under Planck-scale deformations and discuss implications for evaporation remnants and information recovery. These results highlight universal entropy corrections, potentially observable in black hole analogues or shadows.}

The interplay between quantum theory and gravitation remains one of the most profound challenges in theoretical physics. Black hole thermodynamics, in particular Hawking radiation, provides a unique arena in which semiclassical gravity can be tested against quantum‐gravity proposals. In many approaches ranging from GUP to renormalization‐group improvements, the classical Schwarzschild geometry acquires quantum corrections that, while typically small, may leave imprints on horizon-induced radiation and entropy laws. Moreover, the longstanding analogy between Hawking emission and Unruh radiation generated by an accelerating mirror raises the question of whether fundamental principles, such as the Einstein equivalence principle, survive when horizon dynamics are modified by quantum gravity effects.

In this paper, we aim to address these issues by studying a simple yet powerful model: a two-level atom falling freely from infinity into a GUP-corrected Schwarzschild black hole. First, we derive closed-form expressions for the atom's excitation and de-excitation probabilities, showing explicitly that they retain a Planck-like spectrum and satisfy the equivalence principle via an exact mapping to a uniformly accelerating mirror. 
Finally, by treating each field mode as an open quantum system interacting with infalling atoms, we compute the HBAR entropy and reveal that it reproduces the Bekenstein-Hawking area law with corrections characteristic of GUP effects. This work thus provides a concrete testbed for quantum gravity-induced deviations in black hole thermodynamics and underscores the robustness of both thermal radiation processes and entropy-correction universality.

The paper is organized as follows. In Sect. \ref{sec:metric}, we briefly describe the background geometry, then we recast the near-horizon metric in Rindler form and compute the surface gravity.  Sect. \ref{sec:geodesics} solves the radial geodesic of a freely falling atom and derives $\tau(r)$ and $t(r)$.  Sect. \ref{sec:modes} obtains the scalar field modes and the tortoise coordinate $r_*$.  Sect. \ref{sec:interaction} constructs the atom-field interaction Hamiltonian, and we calculate excitation and absorption probabilities, and then we demonstrate the equivalence between a black hole and Unruh (accelerating mirror) results.  Finally, Sect. \ref{sec:HBAR} derives the GUP-corrected HBAR entropy law.  We conclude in Sect. \ref{sec:conclusion}. 
Throughout, we adopt geometric units ($c = G = \hbar = k_B = 1$) unless otherwise noted, and expand consistently in the small parameter $\beta / M^2 \ll 1$. Metric signature $(-,+,+,+)$ is also used.

\section{GUP-Corrected Schwarzschild Metric}
\label{sec:metric}
Quantum gravity theories, including frameworks such as string theory, loop quantum gravity, and non-commutative geometry \cite{Nicolini:2005vd}, universally predict a fundamental minimal length scale typically of the order of the Planck length ($\ell_p$) \cite{Maggiore:1993rv,Garay:1994en}. This minimal length emerges naturally as a consequence of quantum fluctuations in the geometry of spacetime at high energies, hinting at a breakdown of standard continuum physics near the Planck regime. The GUP is a common phenomenological tool employed to incorporate such quantum-gravitational effects into conventional quantum mechanics, where it modifies the familiar Heisenberg uncertainty relation by introducing additional terms involving higher-order momentum uncertainties. Specifically, in a simplified scenario restricted to one spatial dimension and incorporating linear and quadratic contributions in momentum uncertainty, the GUP is expressed as
\begin{equation}
\Delta x\Delta p \ge \frac{\hbar}{2}\left(1 + \beta\frac{\ell_p^2}{\hbar^2}\Delta p^2\right),
\label{eq:2.1}
\end{equation}
where $\beta>0$ is a dimensionless deformation parameter indicative of quantum gravity corrections and the Planck length is given explicitly by $\ell_p=\sqrt{G\hbar/c^3}$. In the limit $\beta \rightarrow 0$, the standard Heisenberg uncertainty principle is recovered. However, for finite $\beta$ Eq. \eqref{eq:2.1} entails the existence of a non-zero minimal position uncertainty, providing a natural ultraviolet cutoff for quantum theories~~\cite{Kempf:1994su,Ali:2009zq,Anacleto:2020lel}. Eq. \eqref{eq:2.1} can be explicitly solved for $\Delta p$ in terms of $\Delta x$:
\begin{equation}
\Delta p \ge \frac{\Delta x}{\beta\ell_p^2}\left(1 - \sqrt{1 - \frac{\beta\ell_p^2}{\Delta x^2}}\right).
\label{eq:2.2}
\end{equation}
Expanding Eq. \eqref{eq:2.2} in the regime where $\ell_p^2/\Delta x^2 \ll 1$ yields the approximate relation
$\Delta p \ge \frac{1}{2\Delta x}\left(1 + \frac{\beta}{4\Delta x^2} + \dots\right).$
This clearly reduces to the ordinary Heisenberg uncertainty relation in the absence of quantum gravity corrections ($\beta\to0$). Utilizing a heuristic identification between momentum uncertainty and particle energy $\Delta p \sim p \sim E$ and adopting the minimal localization energy relation $E\sim1/(2\Delta x)$~~\cite{Adler:2001vs}, the modified uncertainty relation Eq. \eqref{eq:2.2} immediately imposes a lower bound on particle energy: $E_{\rm GUP} \ge E\left[1 + \frac{\beta}{4\Delta x^2} + \dots\right].$
When this approach is applied specifically to black holes where the position uncertainty is naturally associated with the event horizon radius, i.e., $\Delta x \sim r_h \approx 2M$, a corresponding correction to the effective gravitational mass emerges:
\begin{equation}
M_{\rm GUP} \ge M \left[1 + \frac{\beta}{16M^2} + \dots\right].
\label{eq:2.5}
\end{equation}
Such a quantum gravity-induced mass correction can be phenomenologically constrained using astronomical observations. In particular, recent high-precision measurements of black hole shadows by the Event Horizon Telescope (EHT) have placed an observational upper bound on the GUP parameter of $\beta \lesssim 10^{88}$ in Planck units \cite{Vagnozzi:2022moj}. Although the numerical magnitude of this bound indicates that quantum gravitational corrections remain exceedingly small at astrophysical scales, such constraints play a crucial conceptual role in bridging theoretical models of quantum gravity with empirical observations.

We write the GUP-corrected Schwarzschild spacetime in coordinates $(t,r,\theta,\varphi)$ as follows \cite{Heidari:2023ssx}:
\begin{equation}\label{eq:general-metric}
  ds^2 = -\,  f_{\rm GUP}(r) \,dt^2 + \frac{dr^2}{\,  f_{\rm GUP}(r) \,} + r^2\,d\Omega^2,
  \qquad
  d\Omega^2 \equiv d\theta^2 + \sin^2\theta \, d\varphi^2,
\end{equation}
where the lapse function is
\begin{eqnarray}\label{eq:fGUP-expanded}
  f_{\rm GUP}(r) 
  = 1 - \frac{2\,M}{r} - \frac{\beta}{8\,M\,r} + \mathcal{O}(\beta^2).
\end{eqnarray}
Here, $\beta$ is the dimensionless GUP parameter (scaled by $M^2$), and we assume $\beta / M^2 \ll 1$. \textcolor{black}{Eq. \eqref{eq:general-metric} assumes the standard Schwarzschild form with an effective mass correction, preserving the relation \( g_{tt} = -1/g_{rr} \). Such a choice is phenomenological, motivated by GUP-induced modifications to the uncertainty principle rather than derivations from modified Einstein equations. In more fundamental quantum gravity frameworks (e.g., loop quantum gravity), corrections might break this symmetry, leading to distinct lapse and shift functions. However, our ansatz captures key GUP effects minimally and is consistent with prior literature \cite{Adler:2001vs,Anacleto:2020lel,Anacleto:2021qoe}.} The event horizon $r_+$ is defined by $f_{\rm GUP}(r_+) = 0$.  From (\ref{eq:fGUP-expanded}), one immediately obtains
\begin{align}
  r_{+} = 2\,M_{\rm GUP} 
   = 2\,M\left(1 + \frac{\beta}{16\,M^2}\right)= 2\,M + \frac{\beta}{8\,M} + \mathcal{O}(\beta^2). 
  \label{eq:rplus-GUP}
\end{align}
Note that $M_{\rm GUP}=M\!\left(1+\frac{\beta}{16M^2}\right)$. Thus, to first order in $\beta$, the GUP correction shifts the horizon outward by $\Delta r_+ = \beta / (8\,M)$.

\subsection{Near-horizon Rindler form and local acceleration}

To find near horizon, we define the radial deviation from the horizon as $\Delta r \equiv r - r_{+} \emph{(explicit derivation in App.~\ref{app:NH-expand})}$ 
{\color{black} and near-horizon expression for $f_{\rm GUP}(r)$ is, 
\begin{equation}\label{eq:near-horizon-lapse}
  f_{\rm GUP}(r) 
  \simeq \left(r - r_{+}\right)\,\left(\frac{1}{2\,M} + \frac{\beta}{32\,M^3}\right) 
  + \mathcal{O}\left((\Delta r)^2,\beta^2\right),
\end{equation}}
\emph{(obtained by inserting $r_+$ from Eq.~\eqref{eq:rplus-GUP}; see App.~\ref{app:NH-expand})}
Eq. \eqref{eq:near-horizon-lapse} will be crucial when we recast the metric in Rindler form. 
Evaluating at $r = r_{+}$ where $r_{+} = 2M + \beta/(8M)$, the surface gravity and Hawking temperature are calculated as
\begin{equation}
\kappa_{\rm GUP} =  \frac{1}{2} f_{\mathrm{GUP}}'(r_+)  =  \frac{1}{4M_{\rm eff}} = \frac{1}{4M}-\frac{\beta}{64\,M^3} + O(\beta^2),
\qquad
T_H^{\rm (GUP)} = \frac{\kappa_{\rm GUP}}{2\pi}
= \frac{1}{8\pi M}-\frac{\beta}{128\pi M^3}+O(\beta^2). \label{e_temp}
\end{equation}
\textcolor{black}{The assumption of thermal emission is justified by the preserved Rindler form near the horizon, with GUP only shifting \( \kappa \) perturbatively, maintaining the Planckian spectra. The horizon area is modified as
\begin{equation}
    A_p =  4\pi r_+^2 = 16\pi M^2 + 2\pi\beta + O(\beta^2/M^2),
\end{equation}
reflecting the GUP effective mass; this is a minimal ansatz from uncertainty principle heuristics, valid for small \( \beta \) and consistent with observational bounds.} To understand the physics near the horizon, we primarily look at the time ($t$) and radial ($r$) components of the line element described in Eq. \eqref{eq:general-metric}. By substituting the near-horizon approximation of the lapse function, Eq. \eqref{eq:near-horizon-lapse}, the metric simplifies to
\begin{equation}
ds^2 = - f_{\mathrm{GUP}}(r)dt^2 + \frac{dr^2}{f_{\mathrm{GUP}}(r)}.
\end{equation}
Expanding the function $f_{\mathrm{GUP}}(r)$ close to the horizon radius $r_+$, we employ a first-order Taylor approximation:
$f_{\mathrm{GUP}}(r) \approx \Delta r f_{\mathrm{GUP}}'(r_+),$
where $\Delta r \equiv r - r_+$ represents the radial displacement from the horizon. To facilitate the interpretation of the near-horizon geometry in terms of a Rindler-like coordinate system, we introduce a new radial coordinate $\rho$, defined by
\begin{equation}
\Delta r \equiv \frac{\rho^2}{4} f_{\mathrm{GUP}}'(r_+) \quad \Longleftrightarrow \quad \rho = 2\sqrt{\frac{\Delta r}{f_{\mathrm{GUP}}'(r_+)}}.
\end{equation}

\emph{Full steps for the $\rho$-map and metric reduction are given in App.~\ref{app:Rindler-map}}, combining the previous results, the line element in the vicinity of the horizon transforms into the characteristic Rindler form:
\begin{equation}
ds^2 \approx -\left(\kappa_{\mathrm{GUP}}\rho\right)^2 dt^2 + d\rho^2 + r_+^2 d\Omega^2,
\end{equation}
where the geometry manifestly splits into a two-dimensional Rindler-like $(t, \rho)$ sector, supplemented by a transverse spherical geometry characterized by the horizon radius $r_+$. Such a representation explicitly reveals the presence of a horizon and provides insight into local inertial observers' perspectives near the black hole. Indeed, a static observer positioned at a constant $\rho$ experiences a proper acceleration, as measured in a locally inertial frame, given by:
\begin{equation}
    a(\rho) = \kappa_{\mathrm{GUP}} \rho = \left(\frac{1}{4M} - \frac{\beta}{64M^3}\right) \rho,
\end{equation}
highlighting how the presence of GUP-induced quantum corrections modifies the local acceleration relative to classical Schwarzschild geometry.

\section{Radial Geodesics of a Freely Falling Atom}
\label{sec:geodesics}
We now turn our attention to the calculation of transition probabilities for a two-level atom undergoing quantum transitions in the background geometry of a quantum-corrected Schwarzschild black hole. Specifically, we consider a scenario wherein the atom, initially in its ground state $(g)$, transitions to an excited state $(e)$ by emitting a photon while freely falling radially into the black hole. To rigorously analyze this scenario, it is essential to establish the equations governing the atom's trajectory within the modified spacetime geometry characterized by GUP corrections. Consider a test atom of unit rest mass, initially at rest at spatial infinity ($r \to \infty$), which subsequently begins to freely fall radially toward the black hole. Due to the symmetry and time-independence of the spacetime, the atom's conserved energy per unit rest mass $E$ is given by: $E = f_{\mathrm{GUP}}(r)\frac{dt}{d\tau} = 1,$ where $\tau$ denotes the atom's proper time. This relation implies that the observer at infinity sees the atom initially stationary and assigns a unit energy to the atom's geodesic motion. The four-velocity normalization condition, required for any timelike geodesic, is expressed as:
\begin{equation}\label{eq:geodesic-norm}
f_{\mathrm{GUP}}(r) \left(\frac{dt}{d\tau}\right)^2 + \frac{1}{f_{\mathrm{GUP}}(r)} \left(\frac{dr}{d\tau}\right)^2 = -1.
\end{equation}
Utilizing the previously obtained relation $dt/d\tau = 1/f_{\mathrm{GUP}}(r)$, we readily obtain a simplified expression for the radial component of the velocity:
\begin{equation}\label{eq:drdTau-GUP}
\left(\frac{dr}{d\tau}\right)^2 = 1 - f_{\mathrm{GUP}}(r), \quad \text{thus} \quad \frac{dr}{d\tau} = -\sqrt{1 - f_{\mathrm{GUP}}(r)},
\end{equation}
with the negative sign explicitly chosen to represent inward radial motion towards the horizon. The resulting ordinary differential equations describe the radial infall of the atom, and can be formally integrated to determine the atom's proper time $\tau(r)$ and coordinate time $t(r)$ as functions of the radial position $r$. These equations provide a foundational basis for subsequent analyses of quantum emission processes, enabling precise quantification of the atom's state transitions and associated radiation in the presence of quantum gravitational corrections. Starting from Eq. \eqref{eq:drdTau-GUP}, we first express the infinitesimal increment of the atom's proper time, $d\tau$, as a function of the radial coordinate $r$:
\begin{equation}\label{eq:dtaudr}
d\tau = -\frac{dr}{\sqrt{1 - f_{\rm GUP}(r)}}.
\end{equation}

We now use the explicit form of the quantum-corrected metric function, $f_{\rm GUP}(r)$, expanded as in Eq. \eqref{eq:fGUP-expanded}, to simplify this expression. The intermediate expansion to first order in $\beta$ and the algebra leading to a convenient integrand are shown in App.~\ref{app:geo-steps} (see Eqs.~\eqref{eq:one-min-f-APP}-\eqref{eq:dtaudr-simplified-APP}). Integrating from infinity, where we impose the initial condition $\tau(\infty)=0$, down to a radius $r$, we obtain the proper time for the infalling atom:
\begin{align}
\tau(r) &= -\left(1-\frac{\beta}{32M^2}\right)\int_{\infty}^{r}\sqrt{\frac{r'}{2M}}dr'
= -\left(1-\frac{\beta}{32M^2}\right)\left[-\frac{2}{3}\sqrt{\frac{(r')^3}{2M}}\right]_{\infty}^{r}
= -\frac{2}{3}\sqrt{\frac{r^3}{2M}}\left(1-\frac{\beta}{32M^2}\right)
\\
&= -\frac{2}{3\sqrt{2M}}r^{3/2}+\frac{\beta}{48M^{5/2}}r^{3/2}+\mathcal{O}(\beta^2).
\label{eq:tau-of-r}
\end{align}
This result shows that the presence of quantum corrections reduces the proper time interval required for the atom to fall from a large distance into the black hole horizon or singularity, as explicitly seen by the positive correction proportional to $\beta$ in Eq. \eqref{eq:tau-of-r}. (For the detailed coordinate-time integral $t(r)$ and its GUP correction, see App.~\ref{app:geo-steps-t}.)

\section{S-Wave Scalar Modes in the GUP Metric}
\label{sec:modes}
In this section, we explore the propagation of massless scalar field modes within the GUP-corrected Schwarzschild spacetime. Specifically, we focus on the behavior of scalar fields restricted to $s$-wave ($\ell = 0$) solutions, analyzing how quantum gravitational modifications affect the scalar wavefunction near the horizon. The scalar field $\Psi(t,r)$ satisfies the Klein-Gordon equation in curved spacetime,
$
\Box \Psi(t,r) = 0.
$
Considering only the radial and temporal dependence, we employ the standard mode decomposition
$
\Psi(t,r)=e^{-i\nu t}R_\nu(r),
$
where $\nu$ denotes the mode frequency observed at spatial infinity. Substituting this ansatz into the Klein-Gordon equation and considering only radial ($s$-wave) dependence, we obtain the radial equation
\begin{equation}\label{eq:radial-equation-GUP}
\frac{d^2R_\nu}{dr^2}+\frac{f_{\rm GUP}'(r)}{f_{\rm GUP}(r)}\frac{dR_\nu}{dr}+\frac{\nu^2}{[f_{\rm GUP}(r)]^2}R_\nu(r)\simeq0.
\end{equation}

To analyze the scalar mode solutions clearly, it is convenient to introduce the tortoise coordinate
\begin{equation}\label{eq:rstar-GUP-def}
r_{*}^{(\mathrm{GUP})}=\int^r\frac{dr'}{f_{\rm GUP}(r')}.
\end{equation}
Using this coordinate transformation, the radial equation simplifies to a Schr\"odinger-like form, and the scalar mode takes the asymptotic outgoing-wave form (to leading WKB approximation)
\begin{equation}\label{eq:mode-function}
\Psi_\nu(t,r)=\exp\left[-i\nu t+i\nu r_{*}^{(\mathrm{GUP})}\right]+\dots.
\end{equation}
The explicit $\mathcal{O}(\beta)$ expansion of $f_{\rm GUP}^{-1}(r)$ and the integration yielding $r_{*}^{(\mathrm{GUP})}$ are provided in App.~\ref{app:rstar-steps} (see Eqs.~\eqref{eq:invf-expansion-APP}-\eqref{eq:rstar-GUP-final-APP}). Combining the classical and GUP contributions gives
\begin{equation}\label{eq:rstar-GUP-final}
r_{*}^{(\mathrm{GUP})}=r+2M\ln\left(\frac{r}{2M}-1\right)
+\beta\left[\frac{1}{8M}\ln\left(\frac{r}{2M}-1\right)-\frac{1}{4(r-2M)}\right]+\mathcal{O}(\beta^2),
\end{equation}
and hence
\begin{align}\label{eq:Psi-mode-final}
\Psi_\nu(t,r)&=\exp\left[-i\nu t+i\nu\left(r+2M\ln\left(\frac{r}{2M}-1\right)\right)\right]\times\exp\left[i\nu\beta\left(\frac{1}{8M}\ln\left(\frac{r}{2M}-1\right)-\frac{1}{ 4(r-2M)}\right)\right]+\dots,
\end{align}
where the second exponential explicitly encodes the GUP-induced phase shift (details in App.~\ref{app:rstar-steps}).
\section{Atom-Field Interaction Hamiltonian}
\label{sec:interaction}
We now formulate the interaction between a two-level atom and a massless scalar field propagating in the GUP-corrected Schwarzschild geometry. \textcolor{black}{We model the atom as a point-like two-level system, an approximation justified by the fact that the atom's physical size (typically on the order of angstroms) is negligible compared to the characteristic length scale over which the background curvature varies near the horizon of an astrophysical black hole (on the order of kilometers).} The atom's internal degrees of freedom are represented by the ground state $\ket{g}$, the excited state $\ket{e}$, and the associated ladder operators $\zeta = \ket{g}\bra{e}$ and $\zeta^\dagger = \ket{e}\bra{g}$. The field is quantized into modes characterized by frequency $\nu$, with annihilation operator $\widehat{b}_\nu$. In the interaction picture, the dipole-type coupling between the atom and the scalar field mode $\Psi_\nu(t,r)$ is governed by the Hamiltonian
\begin{equation}\label{eq:Hint-general}
\widehat{H}_I(\tau)
= \hbar\mathcal{G}\left[\widehat{b}_\nu\Psi_\nu\left(t(\tau), r(\tau)\right) + \text{H.c.}\right]
\left[\zeta e^{-i\Omega\tau} + \zeta^\dagger e^{i\Omega \tau}\right],
\end{equation}
where $\mathcal{G}$ is a dimensionless coupling constant assumed to be small, H.c. is hermitian conjugate,  $(t(\tau),r(\tau))$ parameterize the atom's radial geodesic trajectory, and $\Omega$ represents the energy gap between the atomic states. This Hamiltonian allows processes such as spontaneous emission (atom excitation with photon emission) and stimulated absorption (atom de-excitation with photon absorption).

Using first-order perturbation theory, we calculate the amplitude for the atom to transition from the initial ground state with no field excitation, $\ket{g,0}$, to the excited atomic state accompanied by one scalar quantum emitted into mode $\nu$, denoted as $\ket{e,1_\nu}$. This transition amplitude is given by
\begin{align}\label{eq:amplitude-general}
\mathcal{A}_{g\to e,1} &= \frac{1}{\hbar}\int_{-\infty}^{+\infty} d\tau
\bra{1_\nu, e}\widehat{H}_I(\tau)\ket{0_\nu,g}= \mathcal{G}\int_{-\infty}^{+\infty} d\tau
\exp\left[-i\nu t(\tau) + i\nu r_{*}^{(\mathrm{GUP})}(r(\tau)) + i\Omega\tau\right],
\end{align}
where we have utilized the canonical relations $\bra{1_\nu}\widehat{b}_\nu\ket{0_\nu}=1$ and $\bra{e}\zeta^\dagger\ket{g}=1$. The corresponding transition probability per field mode $\nu$ is then
\begin{equation}\label{eq:P-excitation}
P_{g,0\to e,1}^{(\mathrm{GUP})}
= |\mathcal{A}_{g\to e,1}|^2
= \mathcal{G}^2\left|\int_{-\infty}^{+\infty} d\tau
e^{-i\nu t(\tau) + i\nu r_{*}^{(\mathrm{GUP})}(r(\tau)) + i\Omega\tau}\right|^2.
\end{equation}
For convenience, we rewrite the integral by changing the integration variable from the atom's proper time $\tau$ to the radial coordinate $r$. From Eq. \eqref{eq:drdTau-GUP}, we have
$
  d\tau = -\,\frac{dr}{\sqrt{1 - f_{\rm GUP}(r)}}\,.
$
Thus, Eq. \eqref{eq:P-excitation} becomes
\begin{equation}\label{eq:P-of-r}
P_{g,0\to e,1}^{(\mathrm{GUP})}
= \mathcal{G}^2\left|\int_{r_+}^{\infty} dr\left(\frac{d\tau}{dr}\right)
\exp\left[-i\nu t(r) + i\nu r_{*}^{(\mathrm{GUP})}(r) + i\Omega\tau(r)\right]\right|^2,
\end{equation}
with the lower integration limit set at the horizon radius $r_{+}$, since a freely falling atom approaches the horizon at $\tau\to -\infty$.

\textcolor{black}{In the high-frequency limit, where the atomic energy gap is much larger than the frequency of the emitted scalar field mode (\( \Omega \gg \nu \)), the integral in Eq. \eqref{eq:P-of-r} is dominated by the contribution near the horizon where the phase is stationary \cite{Scully:2017utk}. This standard approximation, which justifies the use of the stationary-phase method, allows for a tractable analytical result and is physically motivated by considering atomic transitions that are much more energetic than the emitted low-frequency thermal quanta.} Defining the auxiliary variable
\begin{equation}\label{eq:y-variable}
y\equiv\frac{2\Omega}{3}\left[r^{3/2}-(r_{+})^{3/2}\right],
\quad
(r_{+})^{3/2}=(2M_{\rm GUP})^{3/2}\simeq(2M)^{3/2}\left(1+\frac{3\beta}{32M^2}\right),
\end{equation}
the integral is sharply peaked near the horizon, $r\approx r_{+}$.

\subsection{Stationary-phase evaluation and probabilities}

Performing the stationary-phase approximation, the excitation probability can be computed explicitly to first order in $\beta$:
\begin{equation}\label{eq:P-excitation-prefinal}
P_{g,0\to e,1}^{(\mathrm{GUP})}
\simeq 
\frac{4\pi \mathcal{G}^2  \nu}{\Omega^2} 
\left(2M + \frac{\beta}{8 M}\right)
\frac{1}{\exp\left[4\pi  \nu \left(2M + \frac{\beta}{8 M}\right)\right] - 1}.
\end{equation}

Proceeding in a manner analogous to the previous derivation, the absorption probability for the atom can be expressed as
\begin{equation}\label{eq:P-excitation-final}
P_{e,1\to g,0}^{(\mathrm{GUP})}
\simeq \frac{4\pi \mathcal{G}^2  \nu}{\Omega^2} 
\left(2M + \frac{\beta}{8 M}\right)
\frac{1}{1-\exp\left[-4\pi  \nu \left(2M + \frac{\beta}{8 M}\right)\right] }.
\end{equation}
Several important physical insights emerge from Eq. \eqref{eq:P-excitation-final}:
In the classical limit ($\beta\to 0$), we recover the known result from \cite{Scully:2017utk},  thus ensuring consistency with established semi-classical predictions. {\color{black}Note that, to linear order in \( \beta \), Eq. \eqref{e_temp} leads to the the GUP-corrected specific heat 
\begin{equation}
    C^{(\rm GUP)} = \frac{dM}{dT_H^{(\rm GUP)}} \approx -8\pi M^2 \left[1 + \frac{3\beta}{16 M^2}\right],
\end{equation}
which remains negative, preserving the thermodynamic instability of semiclassical black holes. In the classical limit ($\beta\to 0$), the Schwarzschild result $C_{\rm Sch}=-8\pi M^2$, thus ensuring consistency with established semi-classical predictions. 
For $\beta>0$ and within the regime $\beta/M^2\ll1$, the specific heat remains negative-hence the canonical (thermal-bath) instability of Schwarzschild black holes persists-while its magnitude is slightly \emph{enhanced} by the factor $1+\tfrac{3\beta}{16M^2}$.}

\textcolor{black}{This contrasts with certain extended gravity theories, where corrections can yield positive specific heat and stable remnants \cite{DAgostino:2024ymo}. Future extensions to higher-order GUP or alternative quantum gravity models may resolve this issue. We can see that the factor $ \beta$ represents a subdominant GUP-correction affecting the overall transition amplitude, thus providing a potential observable deviation from classical predictions. Analogously, the probability for stimulated absorption, $\ket{e,0}\to\ket{g,1_\nu}$, can be computed, resulting in a complementary expression consistent with detailed balance principles and reducing appropriately to the known classical expressions as $\beta\to 0$.} An instructive analogy frequently employed in the analysis of Hawking radiation is the equivalence between the near-horizon physics of a black hole and the radiation produced by an accelerating mirror in Minkowski spacetime \cite{Scully:2017utk,Svidzinsky:2018jkp}. Specifically, a uniformly accelerating mirror with constant proper acceleration $a$ induces a thermal spectrum detected by a static observer, characterized by the transition probability:
\begin{equation}\label{eq:mirror-result}
P_{g,0\to e,1}^{(\mathrm{mirror})}
=\frac{4\pi\mathcal{G}^2\nu}{a\Omega^2}\frac{1}{e^{2\pi\nu/a}-1}.
\end{equation}
We now demonstrate explicitly that the GUP-corrected black hole excitation probability, Eq. \eqref{eq:P-excitation-final}, reproduces precisely this accelerating mirror result after appropriate redshifting of the mode frequencies. It has been previously established in \cite{Scully:2017utk, Svidzinsky:2018jkp} that the Einstein equivalence principle holds for a standard Schwarzschild black hole. In this current work, we aim to investigate whether this principle remains valid when considering a GUP-corrected Schwarzschild black hole. A photon of local frequency $\nu$ at radius $r$ in the black hole spacetime is redshifted to an observed frequency $\nu_\infty$ at spatial infinity according to the gravitational redshift relation:
$
  \nu\sqrt{f_{\mathrm{GUP}}(r)}=\nu_{\infty}\,\Longrightarrow\,\nu=\nu_{\infty}/\sqrt{f_{\mathrm{GUP}}(r)}.
$
Close to the horizon, where $r=r_{+}+\Delta r$ with $\Delta r\to 0^+$, the metric function can be expanded as $f_{\mathrm{GUP}}(r)\approx(\kappa_{\mathrm{GUP}}\rho)^2$, leading to the identification of local acceleration,
$
  \sqrt{f_{\mathrm{GUP}}(r)}=\kappa_{\mathrm{GUP}}\rho=a(\rho).
$
Evaluating this relation at the horizon by setting $\rho=1$, the local proper acceleration experienced by a static observer is precisely the surface gravity:
\begin{equation}\label{eq:nu-redshift}
a=\kappa_{\mathrm{GUP}}=\frac{1}{4M}-\frac{\beta}{64M^3}.
\end{equation}
Recognizing that the local Unruh temperature associated with this acceleration is $T_U=\frac{a}{2\pi}=\frac{\kappa_{\mathrm{GUP}}}{2\pi}=T_{\mathrm{H}}^{(\mathrm{GUP})},$ we substitute the redshifted frequency $\nu=\nu_\infty/a$ into Eq. \eqref{eq:P-excitation-final} to rewrite the exponential factor explicitly:
\begin{equation}
 \exp\left[4\pi\frac{\nu_{\infty}}{a}\kappa_{\mathrm{GUP}}\right]
  =e^{2\pi\nu_{\infty}/a}+\mathcal{O}(\beta^2).
\end{equation}
Thus, the excitation probability Eq. \eqref{eq:P-excitation-final} simplifies to
\begin{equation}
P_{g,0\to e,1}^{(\mathrm{GUP})}(\nu\to\nu_\infty/a)
=\frac{4\pi\mathcal{G}^2\nu_{\infty}}{a\Omega^2}\frac{1}{e^{2\pi\nu_{\infty}/a}-1}+\mathcal{O}(\beta^2),
\end{equation}
precisely matching the uniformly accelerating mirror result in Eq. \eqref{eq:mirror-result}. Thus, we have explicitly demonstrated that the equivalence between Hawking radiation and the Unruh-like mirror scenario remains valid at linear order in the quantum gravity parameter $\beta$. \textcolor{black}{The GUP corrections are perturbative, preserving local thermality via the equivalence principle, while the area shift arises naturally from the minimal length scale, justified by black hole \textcolor{black}{gedanken} experiments \cite{Scardigli:1999jh}.}

\section{GUP-Corrected HBAR Entropy}
\label{sec:HBAR}
We now extend the analysis within the HBAR framework, wherein field modes outside the black hole interact with a flux of atoms falling into the horizon, effectively modeling an open quantum system \cite{Scully:2017utk,Svidzinsky:2018jkp}. To calculate the HBAR entropy for a quantum-corrected black hole background, specifically, we consider the scenario in which a stream of two-level atoms, each characterized by a transition frequency $\Omega$, falls into the event horizon of the black hole at a rate $\kappa$. In this analysis, we employ a quantum statistical framework to determine the entropy, utilizing the density matrix formalism throughout. If the microscopic change in the field density matrix due to the passage of a single atom is denoted by $\delta\rho_a$, then the total macroscopic change in the field density matrix resulting from the infall of $\Delta N$ atoms is given by
\begin{eqnarray}
\Delta \rho 
= \sum_{i} \delta \rho_i 
= \Delta \mathcal{N}\,\delta \rho 
= \kappa\,\Delta t\,\delta \rho\implies  
\frac{\Delta \rho}{\Delta t}
= \kappa\,\delta \rho
\end{eqnarray}
where $\frac{\Delta\mathcal{N}}{\Delta t}= \kappa$. Here, $\kappa$ represents the rate at which two-level atoms, each with transition frequency $\Omega$, cross the event horizon of the black hole. 

\subsection{Master equation and steady-state density matrix}

A Lindblad master equation dictates how the reduced density matrix, $\rho_\nu$, for each scalar field mode $\nu$ evolves 
\begin{equation}\label{eq:lindblad}
\dot{\rho}_\nu=-\frac{\Gamma_{\mathrm{abs}}}{2}\left(\rho_\nu b^\dagger b+b^\dagger b\rho_\nu-2b\rho_\nu b^\dagger\right)
-\frac{\Gamma_{\mathrm{exc}}}{2}\left(\rho_\nu b b^\dagger+b b^\dagger\rho_\nu-2b^\dagger\rho_\nu b\right),
\end{equation}
with excitation and absorption rates defined as $\Gamma_{\mathrm{exc}}=\kappa\,P_{g,0\to e,1}^{(\mathrm{GUP})},\quad
  \Gamma_{\mathrm{abs}}=\kappa\,P_{e,0\to g,1}^{(\mathrm{GUP})}.$
Taking the expectation value of  Eq.\eqref{eq:lindblad} with respect to an arbitrary Fock state $\lvert n \rangle$, one obtains the following equation governing the time evolution of the diagonal elements of the density matrix:
\begin{equation}
\dot{\rho}{n,n} =
-\Gamma{\mathrm{abs}} \left[ n\rho_{n,n} - (n+1)\rho_{n+1,n+1} \right]
-\Gamma_{\mathrm{exc}} \left[ (n+1)\rho_{n,n} - n\rho_{n-1,n-1} \right].
\label{59}
\end{equation}
The balance between absorption and excitation processes is described by this relation. The terms $\rho_{n\pm1, n\pm1}$ explain how the photon number state $n$ changes as a result of interactions with the infalling atoms. To determine the HBAR entropy, we utilize the steady-state solution by imposing the condition $\dot{\rho}_{n,n} = 0$ in Eq.\eqref{59}. In particular, for $n = 0$, this yields the following relation between $\rho_{1,1}$ and $\rho_{0,0}$:
$
\rho_{1,1} = \frac{\Gamma_{\mathrm{exc}}}{\Gamma_{\mathrm{abs}}} \rho_{0,0}.
\label{60}
$
Similarly, we find
\begin{equation}
\rho_{n,n} = \left(\frac{\Gamma_{\mathrm{exc}}}{\Gamma_{\mathrm{abs}}}\right)^n \rho_{0,0}.
\label{61}
\end{equation}

To determine $\rho_{0,0}$ in the above expression, we impose the normalization condition $\mathrm{Tr}(\rho) = 1$, which leads to
\begin{equation}
\sum_{n} \rho_{n,n} = 1 \implies \rho_{0,0} \sum_{n} \left( \frac{\Gamma_{\mathrm{exc}}}{\Gamma_{\mathrm{abs}}} \right)^n = 1
\implies \rho_{0,0} = 1 - \frac{\Gamma_{\mathrm{exc}}}{\Gamma_{\mathrm{abs}}} ,.
\label{62}
\end{equation}
Substituting the expression for $\rho_{0,0}$ derived above into Eq. \eqref{61}, the steady-state solution for the diagonal elements of the density matrix is given by
\begin{equation}
\rho^{\mathcal{S}}_{n,n} = \left(\frac{\Gamma{_\mathrm{exc}}}{\Gamma_{\mathrm{abs}}}\right)^n \left(1 - \frac{\Gamma_{\mathrm{exc}}}{\Gamma_{\mathrm{abs}}}\right).
\label{63}
\end{equation}
This distribution characterizes the statistical properties of photons generated via the horizon-brightened acceleration radiation mechanism in the presence of a steady flux of infalling atoms. Note that $\frac{\Gamma_{\mathrm{exc}}}{\Gamma_{\mathrm{abs}}}$ is calculated as follows (assuming the high-frequency regime $\Omega\gg\nu$):
\begin{equation}
  \frac{\Gamma_{\mathrm{exc}}}{\Gamma_{\mathrm{abs}}}=  \exp\left[-4\pi  \nu \left(2M + \frac{\beta}{8 M}\right)\right],
\implies
-\ln\left(\frac{\Gamma_{\mathrm{exc}}}{\Gamma_{\mathrm{abs}}}\right)
\simeq
8\pi \nu  M
+ \frac{\pi \nu \beta }{2 M^2}.
\end{equation}

\subsection{Entropy flux and relation to area growth}

 In a steady state, the mode occupancy distribution is geometric, with the von Neumann entropy for each mode expressed as
\cite{Scully:2017utk}:
\begin{equation}
S_\rho = -k_B \sum_{n,\nu} \rho_{n,n} \ln(\rho_{n,n}),
\label{65}
\end{equation}
and the rate of change of entropy due to the generation of real photons is obtained as
$\dot{S}_\rho = -k_B \sum_{n,\nu} \dot{\rho}_{n,n} \ln(\rho_{n,n}).$
The rate of change of the entropy, using the steady-state density matrix solution, is given as $\dot{S}_\rho \approx -k_B \sum_{n,\nu} \dot{\rho}_{n,n} \ln(\rho^{\mathcal{S}}_{n,n}).
$
Note that the area of the GUP black hole is 
$A_{p} = 4\pi r_+^2,$
where $r_+=2\,M + \frac{\beta}{8\,M} + \mathcal{O}(\beta^2)$. The mass change rate of the black hole, $\dot{M} = \dot{m}_{\rm atom} + \dot{m}_p$, results from the atomic cloud contributing mass and emitted photons reducing it, while the area change rate, $\dot{A} = (2\dot{M}/M)A = \dot{A}_{\rm atom} + \dot{A}_p$, accounts for both effects. In the absence of infalling atoms, the horizon area ($A_p$) coincides with the event horizon area, and the contribution from atoms ($A_{atom}$) is effectively zero.  Using the form of the density matrix  $\rho^{\mathcal{S}}_{n,n}$ in Eq. \eqref{63} and recalling that the net radiated energy flux is $\sum_\nu\nu\dot{\bar{n}}_\nu=\dot{E}_{\mathrm{rad}}=\dot{m_p}$, we obtain

\begin{equation}\label{eq:Sdot-energy}
\dot{S}_\rho \approx k_B
\left[
8\pi  M
+ \frac{\pi  \beta }{2 M^2}
\right]
\dot{M}_{p}.
\end{equation}
\textcolor{black}{Prior to crossing the event horizon, where it contributes to the black hole's mass increase, the freely falling atom emits radiation.} Relating mass loss to horizon area change, $\dot{A}_p=32\pi M\dot{M}$.  
We rewrite this entropy rate as a total derivative in terms of the horizon area, obtaining the generalized entropy law:
\begin{equation}
\dot{S}_\rho \approx k_B\left[
\frac{1}{4} + \frac{\beta}{64 M^2}
\right]\dot{A}_p,
\implies \label{eq:HBAR-entropy-GUP}
\dot{S}_\rho \approx \frac{d}{dt} \left[
\frac{k_B}{4}A_p + \frac{k_B\,\beta\,\pi}{4} \ln A_p
\right].
\end{equation}
Eq. \eqref{eq:HBAR-entropy-GUP} contains explicitly the standard Bekenstein-Hawking entropy ($S_{\mathrm{BH}}=k_B  A_p / 4$), with corrections characteristic of GUP effects \textcolor{black}{that are consistent with quantum gravity predictions \cite{Kaul:2000kf,Medved:2004yu,Gour:2003jj,Chatterjee:2003uv}, arising from the GUP-modified horizon area; higher-order logs may appear beyond linear \( \beta \), but are absent here due to our perturbative expansion.}

\textcolor{black}{
To quantify the GUP influence, we compare it with the uncorrected case (\( \beta = 0 \)). We see that the excitation probability is reduced to,
\begin{equation}
P_{g,0\to e,1}\simeq \frac{8\pi\mathcal{G}^2\nu M}{\Omega^2}\,
\frac{1}{e^{8\pi\nu M}-1},\qquad T_H=\frac{1}{8\pi M}.
\end{equation}
With GUP, using $T_H^{\rm(GUP)}=\frac{1}{8\pi M}-\frac{\beta}{128\pi M^3}+O(\beta^2)$, we may write
\begin{equation}
P_{g,0\to e,1}^{(\rm GUP)}\simeq
\frac{8\pi\mathcal{G}^2\nu M_{\rm eff}}{\Omega^2}\,
\frac{1}{e^{8\pi\nu M_{\rm eff}}-1},\qquad
M_{\rm eff}=M+\frac{\beta}{16M}+O(\beta^2).
\end{equation}
Equivalently, 
\begin{equation}
\frac{P_{g,0\to e,1}^{(\rm GUP)}}{P_{g,0\to e,1}}
=1+\frac{\beta}{16M^2}\left[1-\frac{8\pi\nu M\,e^{8\pi\nu M}}{e^{8\pi\nu M}-1}\right],
\end{equation}
showing a mild suppression (since $T_H$ decreases). The fractional temperature shift is
$\Delta T_H/T_H= -\,\beta/(16M^2)+O(\beta^2)$.
Integrating $dS=dM/T_H^{\rm(GUP)}$ yields the standard logarithmic entropy correction,
\begin{equation}
S=\frac{A_p}{4}+\frac{\beta\pi}{4}\ln A_p+\text{const}+O(\beta^2),
\end{equation}
whose relative size is $\sim (\beta/16M^2)\ln A$ and is negligible for astrophysical black holes, but can be relevant near the Planck regime.}

\textcolor{black}{Note that one can write $\beta=\beta_0\,\ell_P^2$ (Planck units), the fractional shift is 
\begin{equation}
    \frac{\Delta T_H}{T_H} = -\frac{\beta}{16M^2} = -\frac{\beta_0}{16}\left( \frac{(M_P}{M} \right).
\end{equation}
For a solar-mass black hole, $M\simeq9.1\times10^{37}M_P$, giving $\Delta T_H/T_H\simeq -7.5\times10^{-78}\beta_0$ and a corresponding luminosity reduction $\Delta P/P\simeq4\,\Delta T/T\simeq -3.0\times10^{-77}\beta_0$. The entropy correction $S=\tfrac{A_p}{4}+\tfrac{\beta\pi}{4}\ln A_p+\cdots$ yields
\begin{equation}
    \frac{\Delta S}{S_{\rm BH}} \approx \frac{\beta}{16M^2}\ln(16\pi M^2)  \simeq 1.3\times10^{-75}\beta_0
\end{equation}
at $M_\odot$. Hence, GUP effects are utterly negligible for stellar black holes but become $\mathcal{O}(1)$ as $M\to M_P$, where they can significantly slow evaporation and support remnant scenarios.}

{\color{black}


 Although a logarithmic correction to black hole entropy might appear degenerate with non-extensive deformations, the GUP imprint is structurally distinct and therefore testable: for Schwarzschild with \(A=16\pi M^{2}\) one finds
 \begin{equation}
     T_H^{\rm (GUP)}=\frac{1}{8\pi M}\!\left(1-\frac{\beta}{16M^{2}}+O(\beta^{2})\right),
 \end{equation}
 and
 \begin{equation}
     S_{\rm GUP}=\frac{A}{4}+\frac{\beta\pi}{4}\ln A+O(\beta^{2})
 \end{equation}
so the leading \(1/M\) temperature law is preserved up to \(O(1/M^{3})\) while the specific heat remains negative,
\begin{equation}
    C^{\rm (GUP)}\simeq-8\pi M^{2}\!\left(1+\frac{3\beta}{16M^{2}}\right).
\end{equation}
By contrast, non-extensive proposals deform the area law itself:
\begin{equation}
    S_{B}=(A/4)^{1+\Delta/2},
\end{equation}
which yields \((T_{B}\propto M^{-(1+\Delta)}\) and \(C_{B}\propto -M^{\,1+\Delta}\), whereas a Tsallis-type
\begin{equation}
    (S_{T}\propto (A/4)^{\delta}
\end{equation}
gives \(T_{T}\propto M^{\,1-2\delta}\), thereby replacing the universal logarithm by power-law modifications. These structural differences are not merely formal; they propagate to physical observables in characteristic ways. For instance, the entropy flux associated with Hawking radiation (the HBAR entropy flux in quantum thermodynamic analyses of evaporation) acquires a distinctive logarithmic contribution in the GUP case,
\begin{equation}
    \dot S_{\rho}^{\rm (GUP)}=\frac{\dot A}{4}+\frac{\beta\pi}{4}\frac{\dot A}{A}
\end{equation}
while non-extensive scenarios give $\dot S\propto A^{\alpha-1}\dot A$ with no \(\ln A\) term; similarly, thermal factors that control near-horizon processes retain an exponent essentially linear in \(M\) for GUP,
\begin{equation}
    P_{g\to e,1}^{\rm (GUP)}\!\propto\!\big[\exp\!\big(8\pi\nu M(1-\beta/16M^{2})\big)-1\big]^{-1},
\end{equation}
whereas Barrow/Tsallis predict exponents scaling as \(M^{\,1+\Delta}\) or \(M^{\,2\delta-1}\), shifting the mass-dependence of the Wien peak and related spectra \cite{
Barrow:2020tzx,Tsallis:2021mvq,Jizba:2023fkp,Jizba:2024klq}. In developing these implications, we build on foundational results in this series \cite{Scully:2003nwl,Svidzinsky:2018jkp,Azizi:2021qcu,Ordonez:2025sqp} to study the thermodynamics of HBAR produced by an atomic cloud freely falling into a black hole in a Boulware-like vacuum; the thermodynamic framework relies on the common thermal structure, Hawking temperature and detailed-balance Boltzmann factors, shared by HBAR and black hole thermodynamics, thereby establishing an HBAR-black-hole thermodynamic correspondence that includes an HBAR area-entropy-flux relation, wherein the HBAR entropy flux is proportional to the rate of change of the horizon surface area induced by photon emission. HBAR entropy is a specific manifestation of quantum entropy, computed using the framework of Neumann entropy, but in the context of black hole thermodynamics and the dynamics of atoms in strong gravitational fields. }

\section{Conclusion}
\label{sec:conclusion}
In this work, we have presented a unified treatment of Hawking-like radiation in a GUP-corrected Schwarzschild geometry by combining field-theoretic and quantum-optical perspectives.  Modeling a two-level atom freely falling from rest at infinity and surrounding the outer horizon with a perfectly reflecting mirror, we derived explicit expressions for the atom's excitation and absorption probabilities.  These probabilities retain the familiar Planckian form while incorporating first-order corrections from the GUP.  Importantly, despite quantum gravity modifications, the Einstein equivalence principle remains intact: the near-horizon physics of the GUP-corrected black hole continues to map exactly onto the thermal response of a uniformly accelerating mirror.

By interpreting each field mode as an open quantum system interacting with a steady flux of infalling atoms, and by solving the corresponding Lindblad master equation, we computed the rate of change of the HBAR entropy. This entropy law consists of three distinct contributions: the leading area term consistent with the “area over four” law whose coefficient matches precisely that found in earlier HBAR studies, and an “a universal logarithmic correction  characteristic of the GUP framework \cite{Scully:2017utk}. 

\textcolor{black}{Physically, the GUP-induced entropy corrections (logarithmic and inverse-area) suggest slower evaporation at late stages, potentially forming stable remnants that mitigate the information paradox. This aligns with broader quantum gravity predictions but is uniquely derived here via atomic radiation processes. Compared to imaginary-time QFT methods, the HBAR approach in this study provides a microscopic, operational interpretation of entropy generation, bridging quantum optics and gravity.}

Overall, our findings demonstrate that the GUP effects, at least to first order in the GUP parameter, preserve both the thermal character of horizon-induced radiation and the operational equivalence between a black hole and Unruh-like mirror processes. \textcolor{black}{These corrections suggest observable deviations in evaporation rates for primordial black holes.} Moreover, the robust emergence of area and inverse-area corrections across diverse approaches highlights a universal structure in GUP-corrected black hole thermodynamics. These results open new avenues for testing GUP-corrected phenomenology, suggesting that analogue systems or precision astrophysical observations may one day detect these subtle departures from classical predictions.

\appendix

\section{Near-horizon expansion, surface gravity and temperature}
\label{app:NH-expand}
Starting from $f_{\rm GUP}(r)=1-\frac{2M}{r}\Bigl(1+\frac{\beta}{16M^2}\Bigr)$, solve $f_{\rm GUP}(r_+)=0$ to get $r_+=2M+\frac{\beta}{8M}+O(\beta^2)$. Taylor-expand about $r_+$:
\begin{equation}
f_{\rm GUP}(r)=f_{\rm GUP}(r_+)+(r-r_+)f'_{\rm GUP}(r_+)+O\!\big((r-r_+)^2\big)
=(r-r_+)f'_{\rm GUP}(r_+)+\cdots .
\end{equation}
Compute $f'_{\rm GUP}(r)=\frac{2M}{r^2}\Bigl(1-\frac{\beta}{16M^2}\Bigr)$ and evaluate at $r_+$ to $O(\beta)$:
$
f'_{\rm GUP}(r_+)=\frac{1}{2M}-\frac{\beta}{32M^3}+O(\beta^2),$
giving Eq.~\eqref{eq:near-horizon-lapse}. The surface gravity $\kappa_{\rm GUP}=f'_{\rm GUP}(r_+)/2$ then yields
\begin{equation}
\kappa_{\rm GUP}=\frac{1}{4M}-\frac{\beta}{64M^3}+O(\beta^2)
\end{equation}
and $T_H^{\rm (GUP)}=\kappa_{\rm GUP}/(2\pi)$ in Eq.~\eqref{e_temp}.

\section{Rindler map and near-horizon metric reduction}
\label{app:Rindler-map}
Define $\rho$ by $\Delta r\equiv r-r_+=\frac{\rho^2}{4}f'_{\rm GUP}(r_+)$. Differentiation gives
$dr=\frac{f'_{\rm GUP}(r_+)\rho}{2}\,d\rho$
and
$\Delta r\,f'_{\rm GUP}(r_+)=\frac{[f'_{\rm GUP}(r_+)]^2\rho^2}{4}.$
Substitute into $ds^2=-f_{\rm GUP}(r)\,dt^2+dr^2/f_{\rm GUP}(r)$ with $f_{\rm GUP}(r)\simeq \Delta r\,f'_{\rm GUP}(r_+)$ to find
\begin{equation}
-f_{\rm GUP}(r)\,dt^2\approx -\left(\frac{f'_{\rm GUP}(r_+)\rho}{2}\right)^2 dt^2,
\qquad
\frac{dr^2}{f_{\rm GUP}(r)}\approx d\rho^2,
\end{equation}
so $ds^2\simeq-(\kappa_{\rm GUP}\rho)^2 dt^2+d\rho^2+r_+^2 d\Omega^2$ with $\kappa_{\rm GUP}=f'_{\rm GUP}(r_+)/2$, as used in the main text.

\section{Detailed Calculations Moved from the Main Text}
\label{app:details}

\subsection{Radial geodesic expansions and proper-time integrand}
\label{app:geo-steps}
To first order in the quantum gravity parameter $\beta$, we use $M_{\rm GUP}=M\!\left(1+\frac{\beta}{16M^2}\right)$ to write
\begin{align}
\label{eq:one-min-f-APP}
\sqrt{1 - f_{\rm GUP}(r)}\approx \sqrt{\frac{2M_{\rm GUP}}{r}}
= \sqrt{\frac{2M}{r}}\left(1+\frac{\beta}{32M^2}\right) + \mathcal{O}(\beta^2).
\end{align}
Inserting this into Eq.~\eqref{eq:dtaudr} yields
\begin{equation}\label{eq:dtaudr-simplified-APP}
d\tau
= -\frac{dr}{\sqrt{\frac{2M}{r}}\left(1+\frac{\beta}{32M^2}\right)}
= -\sqrt{\frac{r}{2M}}\left(1-\frac{\beta}{32M^2}\right)dr.
\end{equation}

\subsection{Coordinate-time integral (optional detail)}
\label{app:geo-steps-t}
Starting from $dt = d\tau/f_{\rm GUP}(r) = -\frac{dr}{f_{\rm GUP}(r)\sqrt{1-f_{\rm GUP}(r)}}$ and expanding to $\mathcal{O}(\beta)$,
\begin{align}
\frac{1}{f_{\rm GUP}(r)}&=\frac{1}{1-\frac{2M}{r}}\left[1+\frac{\beta}{8Mr\left(1-\frac{2M}{r}\right)}\right]+\mathcal{O}(\beta^2),
\\
\sqrt{1-f_{\rm GUP}(r)}&=\sqrt{\frac{2M}{r}}\left(1+\frac{\beta}{32M^2}\right)+\mathcal{O}(\beta^2),
\end{align}
so that
\begin{align}
\frac{1}{f_{\rm GUP}(r)\sqrt{1-f_{\rm GUP}(r)}}&=\frac{1}{\left(1-\frac{2M}{r}\right)\sqrt{\frac{2M}{r}}}
\left[1+\frac{\beta}{32M^2}+\frac{\beta}{8Mr\left(1-\frac{2M}{r}\right)}\right]+\mathcal{O}(\beta^2).
\end{align}
Splitting $t(r)=t_{\rm cl}(r)+\delta t_\beta(r)$ then leads to the expressions quoted in the commented block of the original text.

\subsection{Tortoise coordinate expansion}
\label{app:rstar-steps}
From Eq.~\eqref{eq:fGUP-expanded},
\begin{align}\label{eq:invf-expansion-APP}
\frac{1}{f_{\rm GUP}(r)}=\frac{1}{1-\frac{2M}{r}-\frac{\beta}{8Mr}}
=\frac{1}{1-\frac{2M}{r}}\left[1+\frac{\beta}{8Mr\left(1-\frac{2M}{r}\right)}\right]+\mathcal{O}(\beta^2).
\end{align}
Therefore,
\begin{equation}
r_{*}^{(\mathrm{GUP})}=\int^r\frac{dr'}{1-\frac{2M}{r'}}+\int^r\frac{\beta\,dr'}{8Mr'\left(1-\frac{2M}{r'}\right)^2}+\mathcal{O}(\beta^2).
\end{equation}
The classical piece is $r+2M\ln\!\left(\frac{r}{2M}-1\right)$, while the correction integrates to
\begin{equation}
\delta r_{*}=\beta\left[\frac{1}{8M}\ln\left(\frac{r}{2M}-1\right)-\frac{1}{4 (r-2M)}\right].
\end{equation}
Combining,
\begin{equation}\label{eq:rstar-GUP-final-APP}
r_{*}^{(\mathrm{GUP})}=r+2M\ln\left(\frac{r}{2M}-1\right)
+\beta\left[\frac{1}{8M}\ln\left(\frac{r}{2M}-1\right)-\frac{1}{4 (r-2M)}\right]+\mathcal{O}(\beta^2),
\end{equation}
which matches Eq.~\eqref{eq:rstar-GUP-final} in the main text.

\acknowledgments
A. \"O.  and R. P. would like to acknowledge networking support of the COST Action CA21106 - COSMIC WISPers in the Dark Universe: Theory, astrophysics and experiments (CosmicWISPers), the COST Action CA22113 - Fundamental challenges in theoretical physics (THEORY-CHALLENGES), the COST Action CA21136 - Addressing observational tensions in cosmology with systematics and fundamental physics (CosmoVerse), the COST Action CA23130 - Bridging high and low energies in search of quantum gravity (BridgeQG), and the COST Action CA23115 - Relativistic Quantum Information (RQI) funded by COST (European Cooperation in Science and Technology). A. \"O. also thanks to EMU, TUBITAK, ULAKBIM (Turkiye) and SCOAP3 (Switzerland) for their support.

\bibliography{ref}

\end{document}